\documentclass[prb,preprint,showpacs,preprintnumbers,amsmath,amssymb]{revtex4}

\usepackage{graphicx}
\usepackage{bm}

\newcommand{\be}{\begin{equation}}
\newcommand{\ee}{\end{equation}}
\newcommand{\eq}[1]{Eq.~(\ref{#1})}
\newcommand{\fig}[1]{Fig.~\ref{#1}}
\def\bea{\begin{eqnarray}}
\def\eea{\end{eqnarray}}

\def\vq{{\bf q}}

\def\vk{{\bf k}}

\def\qp{{\bf q}_{\parallel}}

\begin{document}

\title{Close inspection of plasmon excitations in cuprate 
superconductors} 

\author{Andr\'es Greco$^{1,2}$, Hiroyuki Yamase$^{3,4}$, and Mat\'{\i}as Bejas$^{1}$}
\affiliation{
{$^1$}Facultad de Ciencias Exactas, Ingenier\'{\i}a y Agrimensura and
Instituto de F\'{\i}sica Rosario (UNR-CONICET),
Av. Pellegrini 250, 2000 Rosario, Argentina\\	
{$^2$}Max-Planck-Institut f\"ur Festk\"orperforschung, Heisenbergstrasse 1,
D-70569 Stuttgart, Germany\\
{$^3$}National Institute for Materials Science, Tsukuba 305-0047, Japan\\
{$^4$}Department of Condensed Matter Physics, Graduate School of Science, Hokkaido University, Sapporo 060-0810, Japan
}

\date{\today}

\begin{abstract}
Recently resonant inelastic x-ray scattering experiments reported 
fine details of the charge excitations around the in-plane momentum $\qp=(0,0)$   
for various doping rates in electron-doped cuprates ${\rm La_{2-x}Ce_xCuO_4}$. 
We find that those new experimental data are well captured by 
acousticlike plasmon excitations in a microscopic study of the layered 
$t$-$J$ model with the long-range Coulomb interaction. 
The acousticlike plasmon is not a usual plasmon typical to the two-dimensional 
system, but has a small gap proportional to the interlayer hopping $t_z$. 
\end{abstract}

\maketitle

\section{Introduction} 
The x-ray scattering technique is nowadays commonly used 
in cuprate high-$T_c$ superconductors. 
X-ray diffraction and resonant x-ray scattering 
revealed a charge-order tendency 
around the in-plane momentum $\qp \approx (0.6\pi,0)$ and $(0.5\pi,0)$  
in hole-doped cuprates (h-cuprates) \cite{ghiringhelli12, chang12, achkar12} 
and electron-doped cuprates (e-cuprates) \cite{da-silva-neto15,da-silva-neto16}, respectively. 
Resonant inelastic x-ray scattering  (RIXS) 
(Refs.~\onlinecite{ishii05,wslee14,ishii14,ishii17,hepting18,lin20}) 
clarified a V-shaped dispersion of charge excitations around $\qp =(0,0)$, which extends up to a few eV.  

While charge order was extensively discussed in terms of the so-called spin-charge stripes \cite{tranquada95}    
in cuprates \cite{kivelson03,vojta09}, 
those new x-ray experimental data 
were not in line with such a {\it conventional} scenario. 
The momentum of charge order and charge excitations are not correlated with that characterizing spin fluctuations, 
and in fact
no spin order is accompanied by the charge order.
Hence different types of charge orders 
were intensively explored for h-cuprates \cite{bejas12,allais14,meier14,wang14,atkinson15,yamakawa15,mishra15}, 
but the origin of the charge order is still under debate. 
For e-cuprates, a large-$N$ theory of 
the $t$-$J$ model \cite{bejas17} found that charge excitations are characterized 
by a dual structure in energy space. In a low-energy region, typically with a scale less than 
the exchange interaction $J$, various bond-charge excitations are present. Among others, 
bond-charge excitations with a $d$-wave symmetry exhibit a softening around $\vq=(0.5\pi, 0)$ 
(Refs.~\onlinecite{yamase15b,bejas17,yamase19b}), consistent with the experiment 
data \cite{da-silva-neto15,da-silva-neto16}. 
On the other hand, in the high-energy region, typically with a scale larger than $J$, 
plasmon excitations, i.e., collective on-site charge excitations, 
become dominant and explain the charge excitations observed around $\qp=(0,0)$ 
(Refs.~\onlinecite{greco16,bejas17,greco19}).   

In the context of the $t$-$J$ model, both h- and e-cuprates are studied on an equal footing 
by changing the sign of the second nearest-neighbor hopping $t'$ (Refs.~\onlinecite{tohyama94,gooding94}). 
However, the extension of the large-$N$ theory of the $t$-$J$ model to h-cuprates 
cannot capture the observed charge-order tendency around  $\qp \approx (0.6\pi,0)$ (Ref.~\onlinecite{bejas12}). 
While the large-$N$ theory can be formulated in different schemes \cite{affleck89,morse91,wang92,zeyher96,vojta99}, 
this drawback may not come from a large-$N$ scheme employed in Ref.~\onlinecite{bejas12}. 
Rather it may lie in the fact that the charge-order tendency was observed inside the  pseudogap state 
in h-cuprates, but the theoretical calculations \cite{yamase15b,bejas17,yamase19b} 
were performed in a normal metallic state. Such calculations  \cite{yamase15b,bejas17,yamase19b} may be 
a reasonable description only for e-cuprates, where 
the pseudogap is absent or very weak \cite{armitage10}. 

On the other hand, as in the case of e-cuprates, the large-$N$ theory captures 
the high-energy charge excitations observed in h-cuprates \cite{ishii17} in terms of plasmons \cite{greco19}. 
The plasmon excitations, therefore, seem universal in both h- and e-cuprates. 
However, the origin of the high-energy charge excitations are under debate 
and mainly three different scenarios are proposed: 
(i) a certain collective mode near a quantum phase transition, which is specific 
to e-cuprates \cite{wslee14,dellea17}, 
(ii) intraband particle-hole excitations \cite{ishii05,ishii14,ishii17} present in both e- and h-cuprates, 
and (iii) plasmon excitations with finite out-of-plane momentum $q_z$
 \cite{greco16,bejas17,greco19,hepting18}, 
which should be present in both e- and h-cuprates \cite{greco16,bejas17,greco19}.

Recently, Cu $L_3$-edge RIXS experiments reported details of the 
high-energy charge excitations for e-cuprates 
${\rm La_{2-x}Ce_xCuO_4}$ (LCCO) \cite{hepting18,lin20}. 
By using doping-concentration-gradient films, 
the authors in Ref.~\onlinecite{lin20} reported fine details of the charge excitations 
as a function of doping, $q_z$, and $\vq_\parallel$, which offer a stringent test of the plasmon scenario 
advocated in Refs.~\onlinecite{greco16,hepting18,greco19}. 
This test is particularly important because some experiments cast doubt 
on the presence of the plasmons \cite{wslee14, ishii14, ishii17, mitrano18,husain19}. 
We find that those detailed data are well understood in terms of 
acousticlike plasmons obtained in the large-$N$ theory of the layered $t$-$J$ model. 


\section{Model}
It is well known that cuprates are correlated electron systems and a minimal model 
of the ${\rm CuO_2}$ planes is the $t$-$J$ model\cite{fczhang88}. 
To understand the high-energy charge excitations around $\qp=(0,0)$, 
the coupling between the adjacent planes is important as shown in a theoretical study \cite{greco16}, 
where a layered $t$-$J$-$V$ model was employed,   
\bea
&&H = -\sum_{i, j,\sigma} t_{i j}\tilde{c}^\dag_{i\sigma}\tilde{c}_{j\sigma}  \nonumber \\
&& \hspace{5mm} +J \sum_{\langle i,j \rangle} \left( \vec{S}_i \cdot \vec{S}_j - \frac{1}{4} n_i n_j \right)
+\frac{1}{2} \sum_{i,j}  V_{ij} n_i n_j \,.
\label{tJV}  
\eea
Here $\tilde{c}^\dag_{i\sigma}$ and $\tilde{c}_{i\sigma}$ are  
the creation and annihilation operators, respectively, of electrons 
with spin $\sigma(=\uparrow, \downarrow)$  
in the Fock space without any double occupancy, 
$n_i$ is the electron density operator, and $\vec{S}_i$ is the spin operator. 
The indices $i$ and $j$ run over the sites of a three-dimensional lattice. 
The hopping $t_{i j}$ takes the value $t$ $(t')$ between the first (second) nearest-neighbor 
sites on a square lattice, and $t_z$ between the adjacent planes. 
$\langle i,j \rangle$ indicates a pair of nearest-neighbor sites on the square lattice and 
the exchange interaction $J$ is considered only inside the plane because 
the out-of-plane exchange term is much smaller than $J$ (Ref.~\onlinecite{thio88}). 
$V_{ij}$ is the Coulomb repulsion. 

Treating the nondouble occupancy constraint within a large-$N$ approximation \cite{greco16},  
the electronic quasiparticles disperse in momentum space as 
\be
\varepsilon_{\vk} = \varepsilon_{\vk}^{\parallel}  + \varepsilon_{\vk}^{\perp} \, ,
\label{Ek}
\ee
where the in-plane dispersion $\varepsilon_{\vk}^{\parallel}$ and the out-of-plane dispersion 
$\varepsilon_{\vk}^{\perp}$ are given by, respectively,  
\bea
&&\varepsilon_{\vk}^{\parallel} = -2 \left( t \frac{\delta}{2}+\Delta \right) (\cos k_{x}+\cos k_{y}) 
\nonumber \\
&& \hspace{30mm} - 4t' \frac{\delta}{2} \cos k_{x} \cos k_{y} - \mu \,, \label{Epara} \\
&&\varepsilon_{\vk}^{\perp} = 2 t_{z} \frac{\delta}{2} (\cos k_x-\cos k_y)^2 \cos k_{z}  \,. 
\label{Eperp}
\eea
The functional form $ (\cos k_x-\cos k_y)^2$ in $\varepsilon_{\vk}^{\perp}$ 
is frequently invoked for cuprates\cite{andersen95}; see also Fig.~1 and Eqs. (4) and (7) in 
Ref.~\onlinecite{yamase06} for  the interlayer hopping integrals in real space. 
Other forms for $\varepsilon_{\vk}^{\perp}$, 
however, do not change the qualitative features of our results. 
Although the electronic dispersion looks like that in a free electron system, 
the hopping integrals $t$, $t'$, and $t_z$ are renormalized by doping $\delta$ 
because of electron correlation effects.  
In addition, the term $\Delta$ in \eq{Epara}, 
which is proportional to $J$, is the mean-field value of the 
bond variables introduced to decouple the exchange term 
through a  Hubbard-Stratonovich transformation.
The value of $\Delta$ is computed self-consistently together with the chemical
potential $\mu$ for a given $\delta$. 

The term $V_{i j}$ in the Hamiltonian (\ref{tJV}) describes the long-range Coulomb interaction, 
which satisfies Poisson's equation on a lattice. We solve it 
in momentum space and obtain \cite{becca96} 
\be
V(\vq)=\frac{V_c}{A(q_x,q_y) - \cos q_z} \,,
\label{LRC}
\ee
where $V_c= e^2 d(2 \epsilon_{\perp} a^2)^{-1}$ and $A(q_x,q_y)= \alpha (2 - \cos q_x - \cos q_y)+1$ 
with $\alpha=\frac{\tilde{\epsilon}}{(a/d)^2}$ and $\tilde{\epsilon}=\epsilon_\parallel/\epsilon_\perp$; 
$\epsilon_\parallel$ and $\epsilon_\perp$ are the 
dielectric constants parallel and perpendicular to the planes, respectively; 
$a$ and $d$ are the lattice constants in the planes and between the 
planes, respectively; and $e$ is the electric charge of electrons.

In the large-$N$ scheme, the density-density correlation function is renormalized 
already at leading order and can describe collective charge excitations; 
see Ref.~\onlinecite{greco16} for a full formalism of the correlation function. 
We compute the imaginary part of the density-density correlation function as 
a function of $\vq$ and $\omega$ for the parameters 
$t'/t=0.30$, $t_z/t=0.03$, $J/t=0.3$, $V_c/t=8$, and $\alpha=3.2$.   
We consider 30 planes along the $z$ direction to get a reasonable 
resolution of the out-of-plane momentum transfer $q_z$; 
our $q_z$ is given by $q_z=2 \pi n_z /30$ with $n_z$ being integer. 
Along the $x$ and $y$ direction we take the thermodynamic limit. 
The temperature is set zero. 

Our results are compared with the data in Ref.~\onlinecite{lin20}, where 
momentum is given in units of $(2\pi/a,2\pi/b,2\pi/c)$;  $a$, $b$, and $c$ are lattice constants. 
In the present theory, on the other hand, momentum is measured in units of 
$a^{-1}=b^{-1}=d^{-1}=1$. Since LCCO contains two planes in the unit cell, 
the distance between the adjacent planes is $d=c/2$. 
For instance, the momentum $\vq=(0.06,0,1.68)$ 
in Ref.~\onlinecite{lin20} corresponds to $\vq=(0.12, 0, 0.32)\pi$ 
in the first Brillouin zone in our theory.  
We set $t=750$ meV (Ref.~\onlinecite{greco19}) to describe energy 
in units of eV.  

In the present  theory, the long-range Coulomb repulsion and the in-plane electronic correlations are  
treated on an equal footing. Hence our theoretical scheme is different 
from that in Ref.~\onlinecite{hepting18}. 
In Ref.~\onlinecite{hepting18}, the in-plane charge susceptibility 
was calculated first by determinant quantum Monte Carlo in the two-dimensional 
three-band Hubbard model  
and after that the effect of the long-range Coulomb interaction obtained in an layered electron gas model 
was incorporated into the dielectric function. 

\begin{figure}
\centering
\includegraphics[width=7cm]{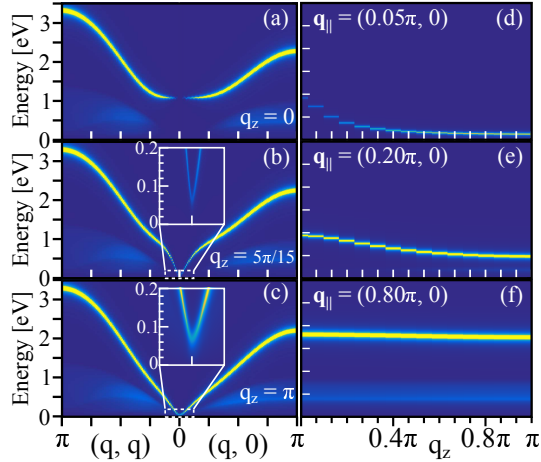}
\caption{Plasmon dispersion as a function of in-plane momentum 
along $(\pi,\pi)$-$(0,0)$-$(\pi,0)$ direction for (a) $q_z=0$, (b) $5 \pi /15$, and (c) $\pi$. 
The doping rate is $\delta=0.17$. 
The broadening is taken as $\Gamma=10^{-2}t$ to make the dispersion sharper. 
Faint spectral weight below 0.8 eV corresponds to the particle-hole continuum and 
becomes broader with increasing $q$. 
The insets in (b) and (c) magnify a very low-energy region around $\qp=(0,0)$. 
Plasmon dispersion as a function of $q_z$ for 
$\vq_\parallel=(0.05,0)\pi$ (d), $(0.2,0)\pi$ (e), 
and $(0.8,0)\pi$ (f).} 
\label{fig4}
\end{figure}

\section{Results and comparison with experiments}  
In Figs.~\ref{fig4}(a)-\ref{fig4}(c), 
we show the plasmon dispersion along $(\pi,\pi)$-$(0,0)$-$(\pi,0)$ direction for 
$q_z=0$, $5 \pi/15$ (close to an experimental value reported in Ref.~\onlinecite{lin20}), 
and $\pi$, respectively. 
The result for $q_z=0$ (upper panel) describes the optical plasmon mode, which 
is in good agreement with the plasmon frequency observed in 
cuprates\cite{singley01,nuecker89,romberg90}. 
Early experiments\cite{uchida91} reported that 
the optical plasmon frequency increases with increasing 
doping, which is well reproduced in the $t$-$J$ model \cite{prelovsek99,greco16}. 
For finite values of $q_z$ (middle and lower panels) 
the plasmon energy at $\qp= (0,0)$ suddenly drops. 
A close look at the results reveals the presence of a gap at $\qp= (0,0)$; 
see the insets in Figs.~\ref{fig4}(b) and \ref{fig4}(c).  
Although the gapless excitations at $\vq_\parallel=(0,0)$, namely acoustic plasmons were discussed 
for finite values of $q_z$ in Refs.~\onlinecite{hepting18} and \onlinecite{lin20} as well as 
early theoretical studies in a layered electron gas model \cite{fetter74,kresin88,bill03}, 
the inter-layer hopping $t_z$ 
yields a finite gap at $\vq_\parallel=(0,0)$ for a finite $q_z$ (Ref.~\onlinecite{greco16}).  
From Figs.~\ref{fig4}(b) and \ref{fig4}(c), we predict 
a gap around $70$ meV for LCCO, i.e., much smaller than 
approximately 300 meV reported for ${\rm Nd_{2-x}Ce_xCuO_4}$  (NCCO) \cite{wslee14}. 
This smaller gap originates mainly from the small value of $t_z/t=0.03$ in the present theory for LCCO 
and seems compatible with the experimental data \cite{hepting18, lin20}. 
Hence we consider that the experimental data are reasonably 
interpreted as {\it acousticlike} plasmons in the sense that they have a small gap at $\qp=(0,0)$. 

Given that the crystal structures of LCCO and NCCO are the same, 
the reason why the value of $t_z/t$ in LCCO can become smaller than NCCO 
needs to be studied further, although new data for NCCO (Ref.~\onlinecite{hepting18}) suggest 
a gap smaller than 300 meV; see the Supplemental Material in Ref.~\onlinecite{hepting18}. 
It is also interesting to explore how the plasmon energy with finite $q_z$ changes by controlling 
the interlayer distance, e.g., by uniaxial pressure or intercalation of some elements between 
the layers, because the present theory predicts that the acousticlike plasmon energy is proportional to 
$t_z$ in a small $t_z$ region \cite{greco16}.


As clarified in Ref.~\onlinecite{greco19}, a characteristic feature of the plasmon excitations 
appears in its substantial $q_z$ dependence for a small in-plane momentum,  
which sharply distinguishes it from the usual intraband particle-hole excitations. 
Figures~\ref{fig4}(d)-\ref{fig4}(f) show the plasmon dispersion as a function of $q_z$ for 
several choices of $\vq_\parallel$. As seen in Figs.~\ref{fig4}(a)-\ref{fig4}(c), the plasmon 
intensity becomes very weak for a small $\qp$, but 
it is discernible in \fig{fig4}(d) that 
the plasmon energy and its intensity show a clear dependence on $q_z$. 
This $q_z$ dependence was reported in  the experiment \cite{hepting18}. 
On the other hand,  both the energy and the intensity become less sensitive to $q_z$ 
for larger values of  $\vq_\parallel$ as shown in Figs.~\ref{fig4}(e) and \ref{fig4}(f). 
Currently no experimental data is available for a larger value of $\qp$.

\begin{figure}[t]
\centering
\includegraphics[width=7cm]{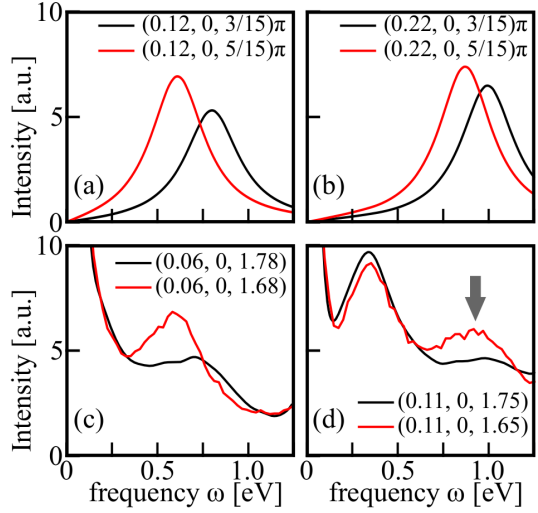}
\caption{ Charge excitation spectra as a function of $\omega$ at $\vq_\parallel=(0.12,0)\pi$ (a) 
and $(0.22,0)\pi$ (b) for $\delta=0.17$ and two different momenta $q_z$. 
(c) and (d) are 
the corresponding experimental data from Figs.~2(c) and 2(d) in Ref.~\onlinecite{lin20}.  
The arrow in (d) indicates excitations of charge origin. 
The peaks at $\omega \approx 0.3$ eV in (d) correspond 
to paramagnon excitations and the peaks at $\omega=0$ in  (c) and (d) are elastic; 
these features are out of a scope of the present work. 
While the in-plane momenta are set exactly the same as the experimental values, 
the values of $q_z$ are slightly different 
between our values and the experimental ones  
because we have a finite number of planes. 
For instance,  $l=1.68$ in the experiment corresponds to $q_z=0.32\pi$ 
which is close to $q_z=5 \pi/15$ in the present calculations for $30$ planes. 
}
\label{fig1}
\end{figure}

To make closer comparisons with the experimental data in Ref.~\onlinecite{lin20}, 
we show in Figs.~\ref{fig1}(a) and (b) charge excitation spectra 
as a function of $\omega$ at $\vq_\parallel=(0.12,0)\pi$  
and $(0.22,0)\pi$ for  $\delta=0.17$; 
experimental values of  $q_z$ are finite and the charge excitation spectrum 
in $\qp$-$\omega$ space corresponds to \fig{fig4}(b), where the acousticlike plasmons are realized.   
A broadening $\Gamma=0.12t$ is introduced phenomenologically 
in the calculations of the charge response. 
The peak energy at both momenta in Figs. 2(a) and 2(b) agrees very well 
with the experimental results Figs. 2(c) and 2(d), 
respectively.  For a given value of $\vq_\parallel$ the experiment shows that the peak energy   
shifts upwards and, at the same time,  the peak intensity decreases with decreasing $q_z$.  
As seen in Figs. 2(a) and 2(b) these two features are reproduced in the present theory. 
In addition, the difference of the peak energy  between the two different values of $q_z$ 
decreases with increasing $\vq_\parallel$ [Figs. 2(c) and 2(d)], which is also captured in
Figs. 2(a) and 2(b).

\begin{figure}
\centering
\includegraphics[width=7cm]{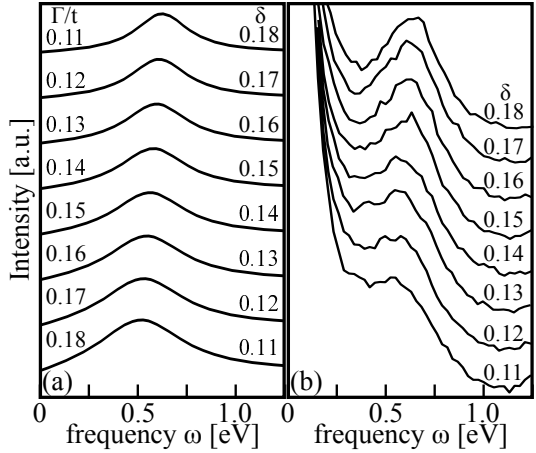}
\caption{ (a) Charge excitation spectra as a function of $\omega$ at $\vq=(0.12,0,5/15)\pi$ for 
different doping rates. The broadening $\Gamma$ is assumed  to decrease with increasing doping. 
(b) The corresponding experimental results from Fig.~4(a) in Ref.~\onlinecite{lin20}. 
Although $q_z$ changes a little by changing doping in the experiment
[see Fig.~4(g) in Ref.~\onlinecite{lin20}], such a change is very small and 
our value of $q_z=5 \pi /15$ is appropriate for all doping. } 
\label{fig2}
\end{figure}

In Fig.~\ref{fig2}(a) we show the doping dependence of the peak energy at $\vq_\parallel=(0.12,0)\pi$. 
With increasing doping  
the peak energy increases monotonically in a way very similar to the experimental results [Fig.~\ref{fig2}(b)]. 
In the experimental results of \fig{fig2}(b) the peak width decreases with increasing doping. 
This implies the suppression of incoherent features with doping. 
To reproduce this feature within the present theory, we have introduced 
a broadening $\Gamma$ (Ref.~\onlinecite{greco19}), which decreases with increasing doping. 
This $\Gamma$ mimics a broadening of the spectrum due to electron correlations obtained 
in a numerical study of the $t$-$J$ model \cite{prelovsek99}. 

\begin{figure}
\centering
\includegraphics[width=7cm]{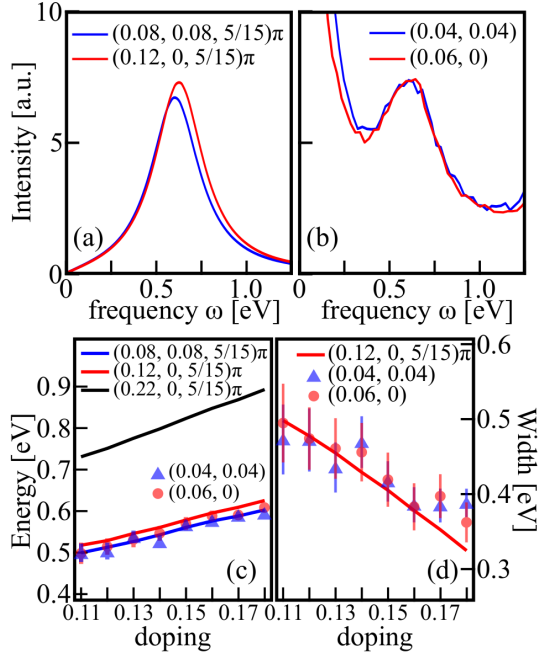}
\caption{  (a) 
Excitation spectra at $\vq=(0.12,0,5/15)\pi$ and 
$(0.08,0.08,5/15)\pi$ for $\delta=0.18$. 
(b) The corresponding experimental data from Fig.~4(d) in Ref.~\onlinecite{lin20}. 
(c) Doping dependence of the peak energy at $\vq=(0.08,0.08,5/15)\pi$ (blue line), 
$(0.12,0,5/15)\pi$ (red line), and $(0.22,0,5/15)\pi$ (black line). 
(d) The peak width at $\vq=(0.12,0,5/15)\pi$ as a function of doping. 
The symbols (triangles and circles) in (c) and (d) are the experimental data 
from Figs.~4(e) and 4(f) in Ref.~\onlinecite{lin20}, respectively.}
\label{fig3}
\end{figure}

Figure~\ref{fig3}(a) shows results at $\vq_\parallel=(0.12,0)\pi$  and  
$(0.08,0.08)\pi$ for doping rate $\delta=0.18$, where 
the peak position practically coincides at those two momenta. 
The peak energy agrees with the experimental data shown in \fig{fig3}(b) and 
the spectral shape of the peak would become essentially the same as the experimental 
data if the background is subtracted properly. 
In \fig{fig3}(c) we show the doping dependence of the peak energy for those 
two momenta (solid lines), which is practically linear in doping and reproduce the experimental results 
(see symbols). While a different doping dependence was discussed in Ref.~\onlinecite{lin20} for a large 
$\qp$ such as $\qp = (0.22, 0)\pi$,  the present theory predicts essentially 
the same behavior also for  $\qp = (0.22, 0)\pi$ [black line in \fig{fig3}(c)] 
at least in a doping region between 0.11 and 0.18. 
In \fig{fig3}(d) we plot the full width at the half-maximum of the plasmon peak 
as a function of doping together with the experimental data, which confirms that 
our phenomenological $\Gamma$ actually works well.  

\section{Outlook}   
We have shown that the acousticlike plasmons obtained in the layered $t$-$J$ model 
explain well the high-energy charge excitation around $\qp=(0,0)$ for e-cuprates. 
For h-cuprates, however, it is not clear whether plasmons can be indeed present. 
The recent momentum-resolved electron energy-loss spectroscopy \cite{mitrano18,husain19} 
cast doubt on the presence of plasmons in
the h-cuprate ${\rm Bi_{2.1}Sr_{1.9}CaCu_2O_{8+x}}$. 
In those papers the authors claimed that 
the dynamical charge response is characterized by 
featureless and momentum-independent excitations. 
On the other hand, Ref.~\onlinecite{ishii17} reported  dispersive 
high-energy charge excitations in h-cuprates ${\rm La_{2-x}(Br,Sr)_xCuO_4}$ by RIXS measurements. 
While the $q_z$ dependence, which is crucial to the plasmon scenario \cite{greco19}, 
was not measured and the obtained dispersion was interpreted as 
incoherent charge excitations in Ref. \onlinecite{ishii17}, 
the observed dispersion as a function of in-plane momentum 
was explained in terms of the acousticlike plasmons \cite{greco19} similar to the present work. 
Our theory therefore implies 
that plasmons exist in both h- and e-cuprates 
in a rather symmetric way \cite{greco19}, 
although many properties such as the pseudogap, superconductivity, and 
antiferromagnetism exhibit a pronounced asymmetry\cite{armitage10} between those systems. 
Further RIXS experiments in h-cuprates are important to clarify whether the high-energy 
charge excitations are indeed plasmons. 

It is natural to ask a possible connection between plasmons and the pseudogap mainly 
observed in h-cuprates. In the present theory, the charge excitations exhibit a dual structure 
in energy space \cite{bejas17}, where plasmon excitations practically 
decouple from the low-energy excitations. 
Hence if low-energy charge fluctuations 
are related to the pseudogap in h-cuprates as frequently discussed in the literature \cite{keimer15,loret19},  
we expect that in contrast to Refs.~\onlinecite{mitrano18} and \onlinecite{husain19},   
high-energy charge excitations 
may not be the major source of the strange metallic properties. 
In fact, high-energy charge excitations extend  
to higher doping and have a doping dependence different from the pseudogap, superconductivity, 
and the strange metallic behavior. 


Plasmons themselves are captured in a layered electron gas model \cite{fetter74,kresin88,bill03}. 
In particular, one might obtain features similar to the present results. 
This does not imply that the electron gas model can be sufficient to 
understand plasmons in cuprates. In fact, as is well accepted, it is not 
an appropriate model of cuprates. 
Moreover, as reported in Refs.~\onlinecite{ghiringhelli12, chang12, achkar12, da-silva-neto15,da-silva-neto16},  
there are also low-energy charge excitations, which seem different from plasmons. 
While the theoretical understanding of Refs.~\onlinecite{ghiringhelli12, chang12, achkar12} is 
still under debate \cite{bejas12,allais14,meier14,wang14,atkinson15,yamakawa15,mishra15},  
we found that the present $t$-$J$ model can capture very well the data presented 
in Refs.~\onlinecite{da-silva-neto15,da-silva-neto16} in terms of $d$-wave bond-charge fluctuations 
\cite{yamase15b,bejas17,yamase19b}. 
This bond-charge physics cannot be captured in the electron gas model. 

In addition, we point out two things which may distinguish the present theory from 
a standard weak coupling study to understand the plasmon physics in cuprates. 
i) The present theory predicts the vanishing of plasmons at half-filling 
whereas they remain even at half-filling in a weak coupling model. 
ii) When we apply the present theory to h-cuprates, 
we expect a doping dependence of the plasmon energy similar to 
\fig{fig3}(c). However, a weak coupling model would predict that 
the plasmon energy decreases with increasing doping. 

\section{Conclusions}  
Recalling that cuprate high-$T_c$ superconductivity is realized by charge carrier doping 
into the Mott insulator, 
the solid understanding of the charge dynamics is definitely indispensable to the cuprate physics. 
In particular, although plasmon excitations are obtained even in an electron gas model, 
it is important to clarify how well a realistic model of cuprates captures the experimental data. 
We have found that the acousticlike plasmons obtained in the layered 
$t$-$J$ model with the long-range Coulomb interaction explain even the fine 
details of charge excitations observed recently 
as a function of doping, in-plane and out-of-plane momenta for Ce-doped ${\rm La_2CuO_4}$.  
The observed charge excitations around $\qp=(0,0)$ are likely due to acousticlike plasmons, 
although some experiments cast doubt on the presence of plasmons. 

\acknowledgments
The authors thank M. Hepting,  P. Horsch, B. Keimer, A. Nag, R. Zeyher, and K.-J. Zhou for very fruitful discussions.  
We thank P. Horsch for a critical reading of the manuscript. 
H. Y. acknowledges support by JSPS KAKENHI Grant No.~JP15K05189 and JP18K18744. 
A. G. acknowledges the Japan Society for the Promotion of Science
for a Short-term Invitational Fellowship program (S17027), under which this work was initiated, 
and Max-Planck Institute for Solid State Research in Stuttgart for hospitality and financial support. 

\bibliography{main} 
\end{document}